\documentclass{article}
\usepackage[latin9]{inputenc}
\usepackage{float}
\usepackage{multirow}
\usepackage{amsmath,bm,amssymb}
\usepackage{graphicx}
\usepackage{hyperref}
\usepackage{ragged2e}
\justifying

\makeatletter

\providecommand{\tabularnewline}{\\}

\usepackage{spconf}
\usepackage{epstopdf}
\usepackage{epsfig}
\usepackage{multicol}
\usepackage{comment}
\usepackage{url}

\title{Promising Accurate Prefix Boosting for sequence-to-sequence ASR}
\name{\parbox{0.9\linewidth}{\center 
Murali Karthick Baskar $^\phi$, Luk\'{a}\v{s} Burget $^\phi$, Shinji Watanabe $^\pi$, Martin Karafi\'{a}t $^\phi$,\\ Takaaki Hori $^\dagger$, Jan ``Honza'' \v{C}ernock\'{y} $^\phi$} \thanks{
The work reported here was carried out during the 2018 Jelinek Memorial Summer Workshop on Speech and Language Technologies, supported by Johns Hopkins University via gifts from Microsoft, Amazon, Google, Facebook, and MERL/Mitsubishi Electric. All the authors from Brno university of Technology was supported by Czech Ministry of Education, Youth and Sports from the National Programme of Sustainability (NPU II) project "IT4Innovations excellence in science - LQ1602" and by the Office of the Director of National Intelligence (ODNI), Intelligence Advanced Research Projects Activity (IARPA) MATERIAL program, via Air Force Research Laboratory (AFRL) contract \# FA8650-17-C-9118. The views and conclusions contained herein are those of the authors and should not be interpreted as necessarily representing the official policies, either expressed or implied, of ODNI, IARPA, AFRL or the U.S. Government. Part of computing hardware was provided by Facebook within the FAIR GPU Partnership Program.
We thank Ruizhi Li, for finding the hyper-parameters to obtain best baseline in WSJ. We also thank Hiroshi Seki, for providing the batch-wise beam search decoding implementation in ESPnet.
}}
\address{
$^\phi$Brno University of Technology, $^\pi$ Johns Hopkins University, \\$^\dagger$Mitsubishi Electric Research Laboratories (MERL)\\
  {\small \tt \{baskar,burget,karafiat,cernocky\}@fit.vutbr.cz,shinjiw@jhu.edu,thori@merl.com}
}

\makeatother

\begin{document}
\ninept \maketitle 
\begin{abstract}
In this paper, we present promising accurate prefix boosting (PAPB), a discriminative training technique for attention based sequence-to-sequence (seq2seq) ASR. PAPB is devised to unify the training and testing scheme in an effective manner. The training procedure involves maximizing the score of each partial correct sequence obtained during beam search compared to other hypotheses. The training objective also includes minimization of token (character) error rate. PAPB shows its efficacy by achieving 10.8\% and 3.8\% WER with and without RNNLM respectively on Wall Street Journal dataset. 
\end{abstract}

\section{Introduction}


Sequence-to-sequence (seq2seq) modeling provides a simple framework to perform complex mapping between input and output sequence. In the original work, where seq2seq~\cite{sutskever2014sequence,bahdanau2014neural} model was applied to machine translation task, the model contains an encoder neural network with recurrent layers, to encode the entire input sequence (i.e the message in a source) into a internal fixed-length vector representation. This vector is an input to the decoder -- another set of recurrent layers with final softmax layer, which, in each recurrent iterations, predicts probabilities for the next symbol of the output sequence (i.e. the message in the target). This work deals with the task of automatic speech recognition (ASR), where the seq2seq model is used to map a sequence of speech features into a sequence of characters. In particular, we use attention based seq2seq model~\cite{bahdanau2014neural}, where the encoder encodes an input sequence into another internal sequence of the same length. The attention mechanism~\cite{bahdanau2014neural}, then focuses on the relevant portion of the internal sequence in order to predict each next output symbol using the decoder. The seq2seq are typically trained to maximize the conditional likelihood (or minimize cross-entropy) of the correct output symbols. For predicting a current character, the previous character (e.g. its one-hot encoding) from ground truth sequence is typically fed as an auxiliary input to decoder during training. This so-called teacher-forcing~\cite{williams1989learning} helped the decoder to learn an internal language model (LM) for the output sequences. During normal decoding, the last predicted character is fed back instead of the unavailable ground truth. Using such training strategy, attention~\cite{chorowski2014end}  based seq2seq model has shown to absorb and jointly learn all the components of a traditional ASR system (i.e. acoustic model, lexicon and language model. Two major drawbacks have been, however, identified with the training strategy described above:
\begin{itemize}
\item \emph{Exposure bias}: The seq2seq training uses teacher forcing, where each output character is conditioned on the previous true character. However during testing, the model needs to rely on its own previous predictions. This mismatch between training and testing leads to poor generalization 
and is referred to as exposure bias~\cite{halsearch,bengio2015scheduled}. 

\item \emph{Error criterion mismatch}: Another potential issue is mismatch in error criterion between training and testing \cite{sak2017recurrent,prabhavalkar2018minimum}. ASR, uses character error rate (CER) or word error rate (WER) to validate the
decoded output while the training objective is the conditional maximum likelihood (cross entropy) maximizing the probability of the correct sequence. 
\end{itemize}

In this work, we first experiment with training objectives that better matches the CER or WER metric, namely minimum Bayes risk (MBR)~\cite{prabhavalkar2018minimum} and softmax margin~\cite{gimpel2010softmax}. We show that such choice of training objective makes teacher-forcing strategy unnecessary and therefore effectively addresses both the aforementioned problems.

Both MBR and softmax margin objective needs to consider alternative sequences (hypotheses) besides the ground truth sequence. Unfortunately, seq2seq model does not make Markov assumptions and the alternative sequences cannot be efficiently represented with a lattice. Instead, we perform beam search to generate an (approximate) N-best list of alternative sequences. However, with the limited capacity of the N-best representation, some of the important hypotheses (i.e. sequences with a low error rate) can be easily pruned out by the beam search, which might result in less effective training. To address this problem, we propose a new training strategy, which we call {\em promising accurate prefix boosting (PAPB)}: The beam search keeps list of $N$ promising prefixes (partial sequences) of the output sequence, which get extended by one character at each decoding iteration. In each iteration, we update parameters of the seq2seq model to boost probabilities of such promising prefixes that are also accurate (i.e. partial sequences with low edit distance to partial ground truth). This is accomplished by using the softmax margin objective (and updates) not only for the whole sequences, but also for all the partial sequence obtained during the decoding.


There are existing works addressing the exposure bias or the error criterion mismatch problem with seq2seq models applied to natural language processing (NLP) problem. For example,  scheduled sampling~\cite{bengio2015scheduled} and SEARN~\cite{halsearch} handle the exposure bias by choosing either the model predictions or the true labels as the feedback to the decoder. The error criterion mismatch is handled using task loss minimization~\cite{bahdanau2015task} using an edit-distance, RNN transducer~\cite{graves2014towards} based expected error rate minimization, and minimum risk criterion based recurrent neural aligner~\cite{sak2017recurrent}.  Few works consider both the problems simultaneously: learning from character sampled from the model using reinforcement learning~\cite{ranzato2015sequence} and actor-critic algorithm~\cite{bahdanau2016actor}. Our work is mostly inspired by beam search optimization (BSO)~\cite{wiseman2016sequence}, where max-margin loss is used as sequence-level objective~\cite{tsochantaridis2005large} for the machine translation task. All the mentioned works were applied to NLP problems, while the focus of this work is ASR. Also, none of the works considered the prefixes (partial sequences) during training. A recent work on seq2seq based ASR was trained with MBR objective~\cite{prabhavalkar2018minimum} using N-best hypotheses obtained from a beam search. However, this work also did not consider the prefixes.
Finally, optimal completion distillation~\cite{sabour2018optimal} technique focuses on prefix learning, but it uses complex learning methods such as policy distillation and imitation learning.

\section{Encoder-Decoder}

With the attention based Encoder-Decoder \cite{chorowski2014end} architecture, the encoder
$H=\mbox{enc}(X)$ neural network provides an internal representations $H=\{h_t\}_{t=1}^T$ of an input sequence $X=\{x_t\}_{t=1}^T$, where $T$ is the number of frames in an utterance. In this work, the encoder is a recurrent network with bi-directional long short-term memory (BLSTM) layers~\cite{hochreiter1997long,schuster1997bidirectional}.
To predict the $l$-th output symbol, the attention component takes the sequence $H$ and the previous hidden state of the decoder $q_{l-1}$ as the input and produces per-frame attention weights
\begin{equation}
\{a_{lt}\}_{t=1}^T=\mbox{Attention}(q_{l-1},H).\label{eq:1-1}
\end{equation}
In this work, we use location aware attention~\cite{bahdanau2016end}. 
The attention weights are expected to have high values for the frames that we need to pay attention to for predicting the current output and are typically normalized to sum-up to one over frames.
Using such weights, the weighted average of the internal sequence $H$ serves as an attention summary vector
\begin{equation}
r_{l}=\sum_t{a_{lt}h_{t}}\label{eq:1-2}.
\end{equation}
The decoder is a recurrent network with LSTM layers, which receives $r_{l}$ along with the previous predicted output character $\overline{y}_{l-1}$  (e.g. its one-hot encodding) as the input and estimates the hidden state vector
\begin{equation}
q_{l}=\mbox{dec}(r_{l},q_{l-1},\overline{y}_{l-1}).\label{eq:2dec}
\end{equation}
This vector is further subject to an affine transformation (LinB) and Softmax non-linearity to obtain the probabilies for the current output symbol $y_{l}$:
\begin{equation}
s_{l}=\mbox{LinB}(q_{l})\label{eq:2dec_1}
\end{equation}
\begin{equation}
p(y_{l}|y_{1:l-1},X)=\mbox{Softmax}(s_{l})\label{eq:2dec_2}
\end{equation}
The probability of a whole sequence $y = \{y_l\}_{l=1}^L$ is 
\begin{equation}
p(y|X)=\prod_{l}^{L}p(y_{l}|y_{1:l-1},X)\label{eq:pyX}
\end{equation}
To decode the output sequence, simple greedy search can be performed, where the most likely symbol is chosen according to~\eqref{eq:2dec_2} in each decoding iteration until the dedicated end-of-sentence symbol is decoded. This procedure, however, does not guarantee in finding the most likely sequence according to~\eqref{eq:pyX}. To find the optimal sequence, multiple hypotheses explored by beam search  usually provides better results.
Note, however, that each partial hypothesis in the beam search has its own hidden state~\eqref{eq:2dec} as it depends on the previously predicted symbols $\overline{y}_{l}$ in that hypothesis.

During training, model parameters are typically updated to minimize cross-entropy (CE) loss for correct output $y^{*}$:
\begin{equation}
\mathcal{L}_{CE}=-\mbox{log}\,p(y^{*}|X) =-\sum_{l=1}\mbox{log} p(y_{l}^{*}|y_{1:l-1}^{*},X).\label{eq:2}
\end{equation}
This is particularly easy with the teacher forcing, when the symbol from the ground truth sequence is always used in~\eqref{eq:2dec} as the previously predicted symbol $\overline{y}_{l-1}$ and, therefore, no alternative hypotheses needs to be considered.

\section{Training criterion}

We compare our proposed PAPB approach with two other objective functions that serves as our baseline. Namely, we use minimum Bayes risk criterion and softmax margin loss, which both perform sequence level discriminative training. Both objectives need to estimate character error rate (CER) for alternative hypotheses, which are explored using beam search. In the following equations, the symbol $\mbox{cer}(y^{*},y)$ denotes the edit distance between the ground truth sequence $y^{*}$ and hypothesized sequence $y$.


\subsection{Minimum Bayes risk (MBR) \label{subsec:Minimum-phone-error}}

In minimum Bayes risk \cite{povey2002minimum,povey2008boosted,vesely2013sequence,su2013error},
the expectation of character error rates $\mbox{cer}(y^{*},y)$ is taken
w.r.t model distribution $p(y|X)$:

\begin{equation}
\mathcal{L}_{MBR}=E_{p({\boldsymbol{y}}|\boldsymbol{X})}\left[\mbox{cer}(y^{*},y)\right]=\sum_{{y}\epsilon Y}p(y|X)\mbox{cer}(y^{*}, y),\label{eq:3}
\end{equation}
In practice, the total set of hypotheses $Y$ generated is reduced to $N$-best hypotheses $Y_{N}$ for computational efficiency.
MBR training objective effectively performs sequence level discriminative
training in ASR \cite{vesely2013sequence} and provides substantial
gains when used as secondary objective after performing cross-entropy
loss based optimization \cite{povey2002minimum}. 

\subsection{Softmax margin loss (SM) \label{subsec:Softmax-margin-loss}}
Softmax margin loss \cite{gimpel2010softmax} falls under the category of maximum-margin \cite{NIPS2011_4184} classification technique. It is a generalization of boosted maximum mutual information (b-MMI)~\cite{povey2008boosted}.
\begin{equation}
\mathcal{L}_{SM}=-s(y^{*},X)\,+\log\left(\sum_{y\epsilon Y} \exp\left(s(y,X)+\alpha \mbox{cer}(y^{*},y)\right)\right), \label{eq:5}
\end{equation}
where $\alpha$ is a tunable margin factor ($\alpha=1$) and the un-normalized score of a chosen sequence,
\vspace{-0.2cm}
\begin{equation}
s(y,X) = \sum_{l}^{L} s_{l}\label{eq:sm}
\end{equation}
is the sum of the pre-softmax outputs $s_{l}$ from~\eqref{eq:2dec_1}. Note the dependence of the scores $s_{l}$ on the chosen hypothesis $y$ through the predictions fed back to decoder in~\eqref{eq:2dec}, which is not explicitly denoted in our notation. 
The function aims to boost the score of the true sequence, $s(y^{*},X)$, to stay above the other hypotheses $s(y,X)$ with a margin defined by CER of the alternative hypotheses.


\section{Promising accurate prefix boosting (PAPB)}

\label{subsec:Prefix-learning}

In PAPB, we perform training at prefix level in similar fashion
to decoding, by incorporating an appropriate training objective with
beam search. The primary motivation to carry out prefix level training is because, seq2seq models predict a sequence, character by character. MBR aims to improves the score of the completed hypothesis with less error, but it might get pruned out during the beam search. However, in our approach, the model will be exposed to all prefixes ${y}_{1:l}$ obtained from N-best as generated by beam search and optimized to maximize the scores of true hypothesis $y^{*}_{1:l}$. In brief, we consider not only the fully completed hypotheses, but also prefixes so that the promising prefixes with low error keep scoring high and therefore are likely to survive the pruning. The $\mathcal{L}_{PAPB}$ loss is computed for each prefix by modifying the softmax margin loss $\mathcal{L}_{SM}$ as:
\begin{equation}
\mathcal{L}_{PAPB}=\sum_{l=1}^{L}\left(-s(y_{1:l}^{*},X)\, \\+\mbox{log}\sum_{{y}\epsilon Y_{N}}\left\{ \exp\left[s({y}_{1:l},X)+B\right]\right\}\right) \label{eq:6}
\end{equation}
where $B=\mbox{cer}(y_{1:l}^{*},{y}_{1:l})$ and $Y_{N}$ denotes the
$N$-best set hypothesis obtained using beam search. 
In equation \eqref{eq:6}, the prefix scores of predicted hypotheses $s({y}_{1:l},X)$ and true hypothesis $s(y_{1:l}^{*},X)$ are computed by summing the scores $s_{l}$ given by \eqref{eq:2dec_1} from $1$ to $l$, while, in the standard sequence objective $\mathcal{L}_{SM}$, the summation is performed only across a whole sequence as noted in~\eqref{eq:sm}.
The contributions of our proposed approach are as follows:
\begin{itemize}
\item The output scores $s({y}_{l},X)$ (as in \eqref{eq:2dec_1})  are computed for each character conditioned on the previous character from the corresponding explored hypothesis (i.e. no teacher-forcing used).
\item In our experiments
, we select the hypothesis from N-best that obtains the lowest CER as the pseudo-true hypothesis $y_{1:l}^{*}$ to compute the score $s(y_{1:l}^{*},X)$, instead of using the true hypothesis. This is to avoid harmful effects during model training by abruptly including the true sequence $y_{1:l}$ into the beam, which might have very small score. Defining the true hypothesis with a pseudo-true hypothesis brings our objective analogous to MBR criterion where very unlikely hypotheses do not affect model parameter updates.
\item The $\mbox{cer}(y^{*}_{1:l},{y}_{1:l})$ is calculated using edit-distance between the prefixes ${y}^{*}_{1:l}$ and \textbf{$y_{1:l}$}. 
Here, the number of characters are kept equal between true prefix $y^{*}_{1:l}$ and prefix hypothesis ${y}_{1:l}$, which, according
to our assumption, should contribute to reduction of insertion and
deletion errors.
\end{itemize}

\section{Experimental setup}

\emph{Database details}: Voxforge-Italian \cite{voxforge} and Wall Street Journal (WSJ) \cite{paul1992design} corpora were used for our experimental analysis. Voxforge-Italian is
a broadband speech corpus (16 hours) and is split into 80\%, 10\% and 10\% to training, development, and evaluation sets by ensuring that no sentence was repeated in any of the sets. WSJ with 284 speakers comprising 37,416 utterances (82 hours of speech) is used
for training, and eval92 test set is used for decoding.

\emph{Training}: Filter-bank features containing 83 dimensional (80 Mel-filter bank
coefficients plus 3 pitch features) coefficients are used as input. In this work, the encoder-decoder model is aligned and trained using attention based approach. Location aware attention~\cite{bahdanau2016end} is used in our experiments. For WSJ experiments, the encoder comprises 3 bi-directional LSTM layers~\cite{schuster1997bidirectional,hochreiter1997long} each with 1024 units and the decoder comprises 2 (uni-directional) LSTM layers with 1024 units. For VoxForge experiments, the encoder comprises 3 bi-directional LSTM layers with 320 units and the decoder contains one LSTM layer with 320 units. The CE training is optimized using AdaDelta \cite{zeilerAdaDelta} optimizer with an initial learning rate set to $1.0$. The training batch size is 30 and the number of training epochs is 20. The learning rate decay is based on the validation performance computed using the character error rate (min. edit distance). ESPnet \cite{watanabe2018espnet} is used to implement and execute all our experiments.
The MBR, softmax margin and prefix training configuration has initial learning rate $0.01$, the number of training epochs is set to 10 and the batch-size to 10. The beam-size for training and testing is set to 10. The model weights are initialized with pre-trained CE model. The rest of configuration is kept the same as for CE training.
In our experiments, we use a modified MBR objective:
\begin{equation}
\mathcal{L}_{MBR}^{'}=\mathcal{L}_{MBR}+\lambda\mathcal{L}_{CE},\label{eq:7}
\end{equation}
which is a weighted combination of the original MBR objective \eqref{eq:3} and CE objective \eqref{eq:7}. Adding the CE objective is analogous to f-smoothing \cite{povey2002minimum}
and provides gains when applied for seq2seq models \cite{prabhavalkar2018minimum}.
Similarly, we also use a modified prefix boosting objective
\begin{equation}
\mathcal{L}_{PAPB}^{'}=\mathcal{L}_{PAPB}+\lambda\mathcal{L}_{CE}\label{eq:8}
\end{equation}
where $\lambda$ is the CE objective weight empirically set to
$0.001$ for both MBR (also noted in \cite{prabhavalkar2018minimum})
and prefix training experiments. Altering the $\lambda$ did not show
much difference in performance.
 
\emph{External language model for WSJ:} Beside the internal language train by the decoder, we have experimented with an external RNN language model (RNNLM) \cite{mikolov2010recurrent} is trained on the text used for seq2seq training along with additional text data accounting to 1.6 million utterances from WSJ corpus. Both character and word-level language models are used in our experiments. The vocabulary size is 50 for character LM and 65k for word LM. The word level RNNLM is trained using 1 recurrent layer of 1000 units, 300 batch size and SGD  optimizer. The character level RNNLM configuration contains 2 recurrent layers of 650 units, 1024 batch size and uses Adam~\cite{DBLP:journals/corr/KingmaB14} optimizer. 

\section{Results and discussion}
We started our initial investigation with Voxforge-Italian dataset and later tested our method on WSJ.
\subsection{Comparison with scheduled sampling (SS)}

 Our best performing CE baseline model is with 50\% SS (which denotes 50 \% true labels) as mentioned in the first row in Table \ref{tab:SS_recognition}. The SS-50\% model is compared
with SS-0\% (0\% true labels) to investigate the impact of only feeding
model predictions. The second row shows that WER of SS-0\% degrades by 8.2 \% on test set and by 5\% on dev set compared to SS-50\% model. 
\begin{table}[t]
\caption{Comparison of recognition performance between scheduled sampling (SS),
MPE and PAPB  on Voxforge-Italian \label{tab:SS_recognition}}
\centering{}{\scriptsize{}{}}%
\begin{tabular}{ccc}
\hline 
{\scriptsize{}{}\%WER }  & {\scriptsize{}{}Test }  & {\scriptsize{}{}Dev}\tabularnewline
\hline 
\hline 
{\scriptsize{}{}SS-50\% (from random init.)}  & {\scriptsize{}{}50.9 }  & {\scriptsize{}{}52.3}\tabularnewline
{\scriptsize{}{}SS-0\% (from random init.)}  & {\scriptsize{}{}59.1 }  & {\scriptsize{}{}57.3}\tabularnewline
{\scriptsize{}{}SS-0\% (fine-tuned from SS-50\%) }  & {\scriptsize{}{}52.9 }  & {\scriptsize{}{}52.5}\tabularnewline
{\scriptsize{}{}MBR }  & {\scriptsize{}{}49.9 }  & {\scriptsize{}{}50.8 }\tabularnewline
{\scriptsize{}{}Softmargin }  & {\scriptsize{}{}50.1 }  & {\scriptsize{}{}50.4 }\tabularnewline
{\scriptsize{}{}PAPB }  & {\scriptsize{}{}47.4 }  & {\scriptsize{}{}47.7}\tabularnewline
\hline 
\end{tabular}
\end{table}
The SS-0\% re-trained using weights initialized from SS-50\% model (acts
as prior), still resulted in performance degradation but the gap got
reduced to 2.0\% on test set and 2.2\% on dev set. These results highlight
the limitation of using scheduled sampling with 0\% true labels (or
100\% model predictions) as it lead to loss in recognition performance.
Thus, a need to use a specific objective which can train only with
model predictions is necessary and justifies the focus of this paper.

\subsection{Comparison of PAPB with MBR}

Table \ref{tab:SS_recognition} shows that the performance of both MBR and softmax margin loss objectives are comparable to each other.
\begin{figure}[t]
\includegraphics[scale=0.32]{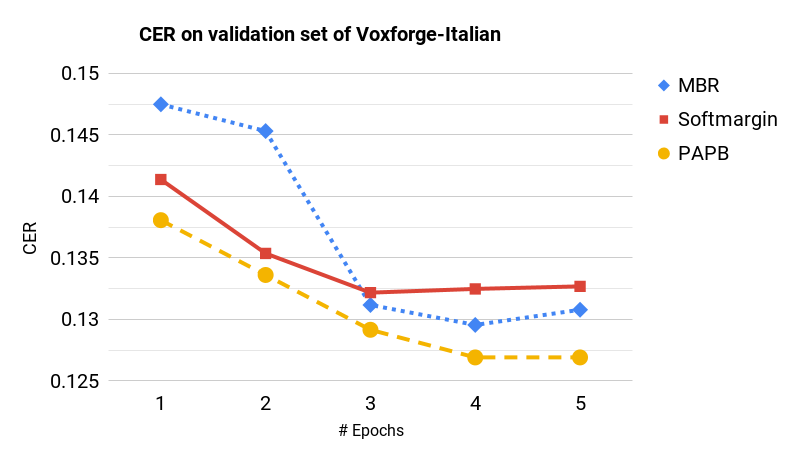}
\caption{Changes in character error rate (CER) during training with different criterion's on dev set of Voxforge-Italian dataset. The plot shows, PAPB improves over both softmax margin loss and MBR objectives}\label{fig:cer}
\end{figure}
While MBR and softmax margin loss provide considerable gains over
scheduled sampling, they do not consider the prefixes (partial
sequences generated during beam search) for training. In the following experiment, we show that the performance of PAPB justify our intuition to use prefix information by providing improvement from 49.9 \% to 47.4 \% WER on test set and from 50.8 \% to 47.7 \% WER on dev set compared to MBR objective. PAPB shows an improvement of 2.7 \% WER for both test and dev sets compared to softmax margin loss. Figure \ref{fig:cer} also shows a similar effect of PAPB noticed during training, by gaining better CER over MBR and softmax margin objectives. 
\begin{table}[t]
\caption{Comparison of recognition performance between different beam sizes
obtained during training $N_{tr}$ and decoding $N_{de}$ for prefix
training on Voxforge-Italian. \label{tab:Beamsize}}
\centering{}{\scriptsize{}}%
\begin{tabular}{cccc}
\hline 
{\scriptsize{}\% WER} & \multicolumn{3}{c}{{\scriptsize{}$N_{de}$}}\tabularnewline
\cline{2-4} 
{\scriptsize{}$N_{tr}$} & \multirow{1}{*}{{\scriptsize{}2}} & \multirow{1}{*}{{\scriptsize{}5}} & \multirow{1}{*}{{\scriptsize{}10}}\tabularnewline
\hline 
{\scriptsize{}2} & {\scriptsize{}50.4} & {\scriptsize{}51.1} & {\scriptsize{}51.5}\tabularnewline
{\scriptsize{}5} & {\scriptsize{}50.8} & {\scriptsize{}49.3} & {\scriptsize{}48.8}\tabularnewline
{\scriptsize{}10} & {\scriptsize{}51.3} & {\scriptsize{}47.9} & \textbf{\scriptsize{}47.4}\tabularnewline
\hline 
\end{tabular}{\scriptsize \par}
\end{table}

\subsection{Effect of varying beam-size during training and testing}
Further analysis on prefix training method is performed to understand the impact of beam-size used during training and testing. The beam-size decides the number of hypotheses to retain during beam search and is denoted as N-best. The results obtained by varying this hyper-parameter showcases the importance of using multiple hypotheses in the loss objective. Table \ref{tab:Beamsize} introduces the effect of retaining best paths (2,5, and 10) during training $N_{tr}$ and testing $N_{de}$. A noticeable pattern observed in our experiments is that increasing the beam-size led to significant improvement in performance. Further, increase in beam-size did not provide considerable gains.

\subsection{Results on WSJ}

\label{sec:Results}

The results in Table \ref{tab:WSJ_results} showcase the importance of using character level, word level RNNLM over no RNNLM.
\begin{table}[t]
\caption{\% CER and \%WER on WSJ corpus for test set with and without LM.
\label{tab:WSJ_results}}
\centering{}{\scriptsize{}}%
\begin{tabular}{cccccccc}
\hline 
{\scriptsize{}LM } & {\scriptsize{}LM } & \multicolumn{2}{c}{{\scriptsize{}CE}} & \multicolumn{2}{c}{{\scriptsize{}MBR}} & \multicolumn{2}{c}{{\scriptsize{}PAPB}}\tabularnewline
\cline{3-8} 
{\scriptsize{}weight} & {\scriptsize{}type} & {\scriptsize{}\%CER} & {\scriptsize{}\%WER} & {\scriptsize{}\%CER} & {\scriptsize{}\%WER} & {\scriptsize{}\%CER} & {\scriptsize{}\%WER}\tabularnewline
\hline 
\hline 
{\scriptsize{}0} & {\scriptsize{}-} & {\scriptsize{}4.6} & {\scriptsize{}12.9} & {\scriptsize{}4.3} & {\scriptsize{}11.5} & \textbf{\scriptsize{}4.0} & \textbf{\scriptsize{}10.8}\tabularnewline
{\scriptsize{}0.1} & {\scriptsize{}char.} & {\scriptsize{}4.6} & {\scriptsize{}11.2} & {\scriptsize{}4.3} & {\scriptsize{}10.1} & {\scriptsize{}4.0} & {\scriptsize{}9.9}\tabularnewline
{\scriptsize{}0.2} & {\scriptsize{}char.} & {\scriptsize{}4.5} & {\scriptsize{}10.9} & {\scriptsize{}4.1} & {\scriptsize{}9.9} & {\scriptsize{}3.9} & {\scriptsize{}9.1}\tabularnewline
{\scriptsize{}1.0} & {\scriptsize{}char.} & {\scriptsize{}2.5} & {\scriptsize{}5.8} & {\scriptsize{}2.5} & {\scriptsize{}5.4} & {\scriptsize{}2.1} & {\scriptsize{}4.5}\tabularnewline
{\scriptsize{}1.0} & {\scriptsize{}word} & {\scriptsize{}2.0} & {\scriptsize{}4.8} & {\scriptsize{}2.1} & {\scriptsize{}4.3} & \textbf{\scriptsize{}2.0} & \textbf{\scriptsize{}3.8}\tabularnewline
\hline 
\end{tabular}{\scriptsize \par}
\end{table}
For decoding with RNNLM, we use look-ahead word LM decoding procedure recently introduced in~\cite{hori2018end} to integrate the word based RNNLM and the character RNNLM is decoded by following the procedure in~\cite{watanabe2017hybrid}. The LM weight is optimized to show the impact of language model across CE, MBR and our proposed PAPB models. The Table \ref{tab:WSJ_results} also show that results of PAPB and MBR shows complementary effect with both word and character LM.

\emph{State-of-the-art results on WSJ}:   Without using external LM deep CNN~\cite{zhang2017very} achieves 10.5\% WER and 9.6\% WER using OCD~\cite{sabour2018optimal}. OCD's nice performance is the resultant effect of their strong baseline with 10.6\% WER. 4.1\% WER are obtained with word RNNLM using end-to-end LF-MMI~\cite{hadian2018end}.

\section{Conclusion}

In this paper, we proposed PAPB, a strategy to train on $N$-best partial sequences generated using beam search. This method suggests that improving the hypothesis at prefix level can attain better model predictions for refining the feed back to predict next character. The softmax margin loss function is inherited
in our approach to serve this purpose. The experimental results shows the efficacy of the proposed approach compared to CE and MBR objectives with consistent gains across two datasets. The PAPB also has its drawbacks in-terms of time complexity, as it consumes 20\% more training time compared to CE training. This work can be further extended to use complete set of lattices instead of N-best list by exploiting the capabilities of GPU for improving time complexity. Also, modified MBR training objective in-place of softmax margin objective can be used to learn prefixes.

\bibliographystyle{ieeetr}
\bibliography{ref_new}

\end{document}